\documentclass[prl,amssymb,amsfonts,twocolumn]{revtex4}
\usepackage{epsfig}
\usepackage{bm}
\usepackage{amsmath}
\usepackage{amsfonts}
\usepackage{amssymb}
\def\beq{\begin{equation}}
\def\eeq{\end{equation}}
\def\bea{\begin{eqnarray}}
\def\eea{\end{eqnarray}}
\def\half{\mbox{$1\over2$}}

\def\bk{{\bf k}}

\def\6{\langle }
\def\9{\rangle }

\begin{document}

\title{ Entangled Photon Pairs from Semiconductor Quantum Dots.}
\author{N. Akopian}
\author{ N. H. Lindner}
\author{ E. Poem}
\author{ Y. Berlatzky}
\author{ J. Avron}
\author{ D. Gershoni}
\email{dg@physics.technion.ac.il}
\affiliation{Department of
Physics, Technion---Israel Institute of Technology, 32000 Haifa,
Israel}
\author{ B. D. Gerardot}
\author{ P. M. Petroff}
\affiliation{Materials Department, University of California Santa
Barbara, CA, 93106, USA }

\begin{abstract}

Tomographic analysis demonstrates that the polarization state of
pairs of photons emitted from a biexciton decay cascade becomes
entangled when spectral filtering is applied. The measured density
matrix of the photon pair satisfies the Peres criterion for
entanglement by more than 3 standard deviations of the
experimental uncertainty and violates Bell's inequality. We show
that the spectral filtering erases the ``which path'' information
contained in the photons color and that the remanent information
in the quantum dot degrees of freedom is negligible.
\end{abstract}


\maketitle

Entanglement, the intriguing correlations of quantum systems
\cite{ref1,bell,ref3} is an essential resource of quantum
information and communication \cite{ref4,ref5,ref6}. Entangled
photons are particularly attractive for applications due to their
non interacting nature and the ease by which they can be
manipulated. Polarization entangled photons are routinely produced
by nonlinear optical effects \cite{ref7,kwiat,ref9}, predominantly
by parametric down conversion \cite{ref7,kwiat}. Such sources have
a large random component whereas quantum data processing schemes
require non-random, or ``event ready'', entangled photons.

Semiconductor quantum dots (SCQDs)
\cite{ref9,ref10,ref11,ref12,ref13,ref14} provide optically
\cite{ref14,ref15} and electrically \cite{ref16} driven sources of
single photons ``on demand''. Compatibility with modern
electronics makes them potential building blocks for quantum
information processing in general \cite{ref17} and sources for
``event-ready'' entangled photons \cite{ref18,ref19} in
particular.

A SCQD biexciton decays radiatively through two intermediate
optically active exciton states \cite{ref20,ref21}. The proposal
that the biexciton-radiative cascade could provide a source of
event-ready polarization entangled photon pairs was made by Benson
et.\ al.\ \cite{ref18}. Entanglement requires two decay paths with
different polarizations, but otherwise indistinguishable. This is
the case if the intermediate exciton states are energetically
degenerate and if, in addition, the final state of the SCQD is
independent of the decay path.

\begin{figure}[htbp]
\epsfxsize=.46\textwidth \centerline{\epsffile{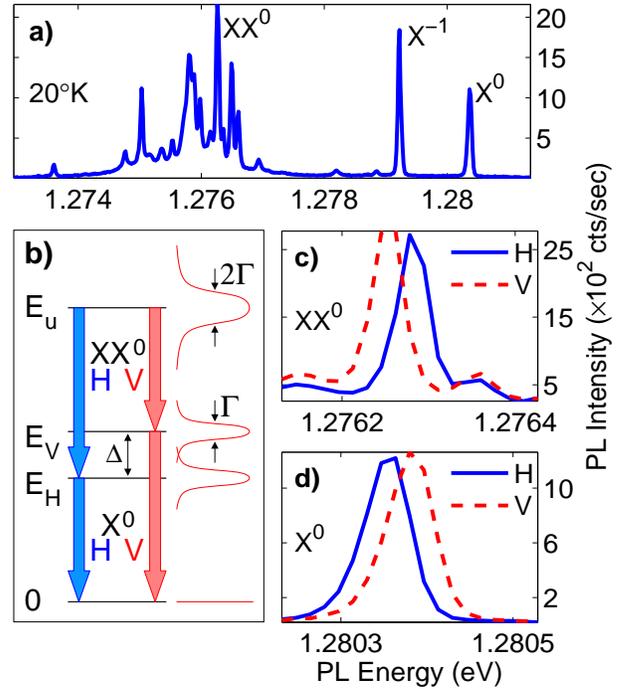}}
\vspace*{-0.2cm}
 \caption{a) PL spectrum of a single SCQD embedded in planar MC.
  (b) Schematic description of the biexciton radiative cascade in natural SCQDs with its
   two co-linearly polarized photons (either H or V), with
energetically distinguishable paths. (c), ((d)) High resolution
polarization sensitive PL spectra of the biexciton XX$^0$ (exciton
X$^0$) lines. The measured spectral width is much larger than the
measured radiative width shown schematically in (b). }
 \label{fig:one}
\end{figure}

The first requirement is difficult to fulfil since the
intermediate exciton states are normally split (by the
electron-hole exchange interaction) \cite{ref25,ref26}. The two
decay paths,  which we denote by horizontal (H) and vertical (V),
have corresponding photon polarizations relative to the asymmetry
axis of the QD (shown in Fig.~\ref{fig:one}). The paths are then
spectrally distinguishable and the polarization state of the
photons cannot be entangled \cite{ref22}. The second requirement,
that the SCQD final state does not depend on the decay path, has
never been tested experimentally.

In this Letter, we show for the first time \cite{nature} that the
``which path'' information can be erased by filtering the photons
spectrally and that this procedure produces an entangled
polarization state. This also proves that the remnant ``which
path'' information residing in the myriad degrees of freedom of
the SCQD, must be small.



We denote by  $|HH\9$ ($|VV\9$) the two photons' polarization
state associated with the H (V) decay path. The final state of the
system is given by
\begin{equation}
  \label{eq:final_state}
  |\psi\9 = \alpha|p_H\9|HH\9|d_H\9 +  \beta|p_V\9|VV\9|d_V\9.
\end{equation}
Here $\alpha$ and $\beta$ are the amplitudes of the decay paths
($|\alpha|^2+|\beta|^2=1$).  The photons' wave packet is denoted
$|p_{H(V)}\9$  and the final states of the SCQD by  $|d_{H(V)}\9$.
All may, a-priori, depend on the decay path.

In Eq.(\ref{eq:final_state}) we neglected the broken paths,
$|HV\9$ and $|VH\9$, which have negligible amplitudes
\cite{ref28}.

The density matrix for the two photons polarization state is given
by tracing out the $|p\9$  and $|d\9$ degrees of freedom:
\begin{equation}
  \label{eq:density_matrix}
  \rho=
  \left(
  \begin{array}{cccc}
  |\alpha|^2 &  0 & 0 & \gamma \\
  0 & 0 & 0 & 0 \\
  0 & 0 & 0 & 0 \\
  \gamma^\star & 0 & 0 & |\beta|^2 \\
  \end{array}
  \right),\ \gamma=\alpha\beta^\star\6 p_H | p_V \9\6 d_H |
  d_V\9\,.
\end{equation}
 The phase of $\gamma$ is a gauge
dependent quantity.

The Peres criterion \cite{peres} applied to
Eq.(\ref{eq:density_matrix}) says that the polarization is
entangled provided $\gamma\neq 0$. This may be re-formulated in
terms of the reliability of the ``which path'' indicator: The
indicator is reliable if its interrogation selects the path, H or
V, with no error. This is the case if $|p_H\9|d_H\9$ is orthogonal
to $|p_V\9|d_V\9$. In this case $\gamma=0$ and the state is
correlated (there are no HV events) but not entangled. If
$\gamma\neq0$, interrogating the indicator does not select a
unique path and the state is entangled. When $|p_H\9|d_H\9$ is
parallel to $|p_V\9|d_V\9$, maximal entanglement, with
$|\gamma|=\half$ , is obtained if the two paths have equal weights
$|\alpha|^2=|\beta|^2=\half$.

Entanglement need not imply violation of Bell's inequality.
Horodecki et al. \cite{ref29} show how to choose the Bell operator
$B$ for a given $\rho$ in order to maximally violate the (CHSH)
Bell inequality. Applying their result to $\rho$ yields
\begin{equation}
  \label{eq:bell_inequality}
  \mbox{Tr}(B\rho)=2\sqrt{1+4|\gamma|^2}.
\end{equation}
Bell inequality is then violated for all $\gamma\neq 0$. Maximal
violation of $2\sqrt{2}$ is obtained for a maximally entangled
state with $|\gamma|=\half$.

Suppose $\6 p_H | p_V \9=0$. Then $\gamma=0$ and the state is not
entangled. Nevertheless, one can entangle such a state by applying
a projection $P$ on the wave packet. This replaces $|\psi\9$ of
Eq.~\eqref{eq:final_state} by $P|\psi\9/{|P|\psi\9|}$ and gives
\begin{equation}
  \label{eq:gamma_tag}
  \gamma'=\frac{\alpha\beta^\star \6 p_H |P| p_V \9}{|P|\psi\9|^2}\6 d_H|d_V \9.
\end{equation}
An optimal choice is such that $\alpha^\star  P|p_H\9=\beta^\star
P|p_V\9$ giving in turn $\gamma'=\half\6 d_H | d_V\9$. Maximal
entanglement is obtained when the final state of the dot $|d\9$
does not depend on the decay path. Thus, $|\gamma'|$ can be
significant, even if $\gamma$  is negligibly small. If the final
state of the SCQD can reliably distinguish between the two
possible decay paths (for example,
due to involvement of different phonons or spins in the radiative
cascade) no entanglement will arise.

Calculating $|p\9$ in second-order perturbation theory within the
dipole and rotating wave approximation \cite{ref30} gives
\setlength\arraycolsep{2pt}
\begin{eqnarray}
\label{eq:amplitudes} A_H
\equiv \alpha\6 \bk_1,\bk_2 |
p_H\9=\frac{e^{i\phi_H}\,\Gamma/2\pi}
{(|\bk_1|+|\bk_2|-\epsilon_u)(|\bk_2|-\epsilon_H)}\,,
\end{eqnarray}
with a similar expression for $A_V$. The index $u$ denotes the
initial biexciton state. The momentum of the photons is labelled
$\bk_1$, $\bk_2$  (using units in which $\hbar=c=1$), and
$\epsilon_j=E_j-\frac{i}{2}\Gamma_j$, $ j=u,\,H,\,V$  are complex
energies with radiative widths  $\Gamma_j$. We implement $P$ as a
projection of the second photon on a rectangular window function
$W(\bk_2)$ of width $w$, centered at the average intermediate
energy $(E_V+E_H)/2$. Plugging this into Eqs.~(\ref{eq:gamma_tag},
\ref{eq:amplitudes})  gives
\begin{equation}
  \label{eq:window_gamma_tag}\gamma'=\frac{\int\!\!\!\int \! d\bk_1 d\bk_2 A_H^\star W
A_V}{\int\!\!\!\int \! d\bk_1 d\bk_2 A_H^\star W
A_H+\int\!\!\!\int \! d\bk_1 d\bk_2 A_V^\star W A_V}\,.
\end{equation}
The denominator, $|P|\psi\9|^2$, is the detection probability of
the two-photons within the spectral window $W$.

From Eq.~\eqref{eq:window_gamma_tag} we calculate $\gamma'$ as a
function of $w$ with $\Gamma_u=2\Gamma_{H/V}$ and the detuning
$\Delta=E_H-E_V$ taken from our measurement. Although one can
easily calculate $\gamma'$ exactly (see Fig.4)
the result is more transparent when one takes into account that
the radiative width $\Gamma$ is the smallest energy scale in the
problem. When the window $w$ is smaller than the detuning,
$\Delta$, $\gamma'$ is actually independent of the radiative width
\begin{equation}
\label{eq:second_photon_filter_gamma_tag}
\left|\gamma'\right|=\frac 1 4\,\left(\frac 1 x
-x\right)\log(\frac{1+x}{1-x}),\quad x=\frac w\Delta\,.
\end{equation}
$|\gamma'|$ reaches the maximal value of $\half$ when the window
is vanishingly small. However, the detection probability
$|P|\psi\9|^2$ vanishes in this limit as well. In the opposite
limit of no filtering, when the window is large, the detection
probability $|P|\psi\9|^2$ approaches unity, while $|\gamma'|$
approaches the residual (small) value $\Gamma/2\Delta$. The
calculation applies a single spectral window for the second photon
(energy conservation is then sufficient to spectrally project the
first photon).

For the measurements we used planar microcavity (MC)
embedded SCQDs samples. The samples were grown by molecular beam
epitaxy on a (100) oriented GaAs substrate. One layer of
strain-induced InAs QDs was deposited in the center of a one
wavelength GaAs microcavity formed by two unequal stacks of
alternating quarter wavelength layers of AlAs and GaAs,
respectively. The height and composition of the QDs were
controlled by partially covering the InAs QDs with a 3 nm layer of
GaAs and subsequent growth interruption. To improve photon
collection efficiency, the microcavity was designed to have a
cavity mode, which matches the QD emission due to ground state e-h
pair recombinations. The sample was not patterned laterally to
prevent obscuration of the emitted photon polarizations.

We used a diffraction limited low temperature confocal optical
microscope for the photoluminescence (PL) studies of single MCQDs
\cite{ref11,ref21}. Temporal correlations between emitted photon
pairs were measured using a wavelength and polarization selective
Hanbury-Brown-Twiss (HBT) arrangement \cite{ref21}. We used a 1
meter monochromator in each arm of the HBT setup to obtain
spectral resolution of $\sim 15\mu$eV. The polarization state of
the emitted light was monitored by the use of liquid crystal
variable retarders (LCVRs) and high quality polarizers.

In Fig.~\ref{fig:one}a we present PL spectrum of a single,
resonant MCQD. The MCQD was excited by a continuous-wave HeNe
laser. The spectrum is composed of sharp lines with linewidths of
roughly 50 $\mu ev$ due to inhomogeneous broadening caused by
temporal changes in the SCQD electrostatic environment. We
identified most of the observed spectral lines using power and
energy dependence polarization sensitive magneto spectroscopy.
Here, we are only interested in the neutral single exciton line
(X$^0$) and the neutral biexciton line (XX$^0$). In Fig 1c and 1d
we present high resolution polarization sensitive PL spectra of
the lines X$^0$ and XX$^0$. Fig.~\ref{fig:one} demonstrates that
the neutral spectral lines XX$^0$ and X$^0$ are composed of two
cross linearly polarized split doublets with detuning of
\mbox{$\Delta = 27 \pm 3 \mu$eV}.
\begin{figure}[htbp]
\epsfxsize=.46\textwidth \centerline{\epsffile{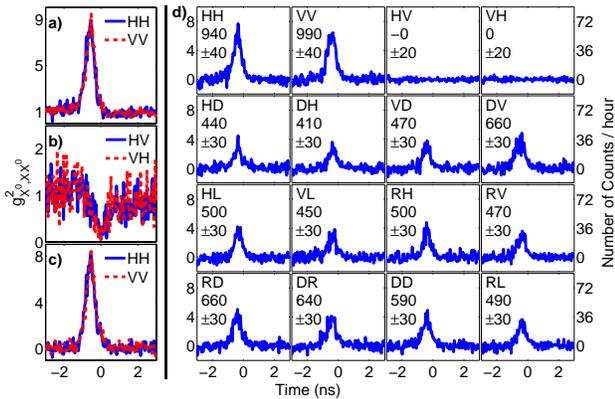}}
\vspace*{-0.2cm}
 \caption{a) (b)) The temporal intensity cross correlation
measurements between the exciton X$^0$ and the biexciton XX$^0$
spectral lines for co- (cross-)linear polarizations. c) The
'reduced' intensity correlation measurements obtained by the
difference between the collinearly polarized curves in a) and the
average curve of the cross-linearly polarized measurements in b).
d) Tomographical measurements of the 'reduced' intensity
cross-correlation functions with spectral resolution of 25
$\mu$eV. D stands for linear polarizer at $45^o$ relative to the H
direction and  R(L) stands for right (left) hand circular
polarizer. The integrated numbers of coincidences in each
measurement are indicated.}
 \label{fig:two}
\end{figure}
We used the HBT setup to measure polarization sensitive temporal
intensity correlations between photons emitted from all the
observed spectral lines. The auto-correlation measurements of the
lines (not shown), show a deep anti-bunching notch at coinciding
times (t=0), demonstrating that each line is a spectral source of
single photons \cite{ref14,ref15,ref16}. The intensity cross
correlation measurements between the neutral exciton X$^0$ and the
neutral biexciton XX$^0$ lines are presented in
Fig.~\ref{fig:two}a~(\ref{fig:two}b) for two different co-
(cross-) linear polarization arrangements. For these measurements,
the excitation intensity was tuned such that both lines were
essentially equal in strength.
When the two polarizers are co-linearly oriented along the major
QD axes (HH or VV) an asymmetric trace is obtained in which the
positive temporal part shows an anti-bunching notch, while the
negative part shows a strong, enhanced bunching peak. This
asymmetrical shape, an experimental signature of an optical
cascade, reveals the temporal sequence of these events. While
emission of a horizontally (vertically) polarized XX$^0$ photon is
followed by emission of a horizontally (vertically) polarized
X$^0$ photon, the opposite never happens \cite{ref18,ref19}. When
the polarizers are cross linearly polarized the bunching trace at
negative times, is replaced by an anti-bunching-like trace. This
is exactly as anticipated by the considerations of
Fig.~\ref{fig:one}b. The linear polarization states of the two
photons emitted during the biexciton-exciton recombination cascade
are completely correlated and both are collinearly polarized.

From the temporal correlation measurement we obtain the radiative
lifetime of the exciton \cite{ref21} $T_{X^0}=0.8 \pm 0.2$~nsec,
which yields $\Gamma=1.6\pm 0.4 \mu$eV  for the excitonic
radiative width.

In Fig.~\ref{fig:two}d) we present the ``reduced'' intensity
correlation functions, obtained by subtracting the average of the
cross-linearly polarized traces from the traces of all the
tomographic correlation measurements. By integrating the
``reduced'' functions over time we obtain the net number of
cascaded coincidences, used for the tomography.
\begin{figure}[htbp]
\epsfxsize=.46\textwidth \centerline{\epsffile{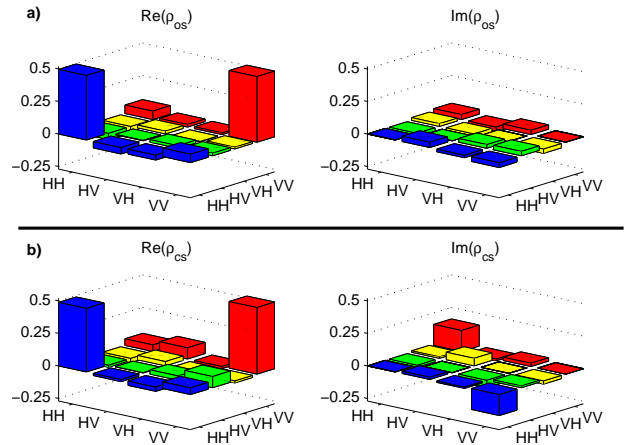}}
\vspace*{-0.2cm}  \caption{The measured two photons' density
matrix: a) (b)) obtained with spectral window of 200 (25)
$\mu$eV.} \label{fig:three}
\end{figure}

We performed two sets of independent measurements of the reduced
intensity cross correlation function in 16 different combinations
of the polarizers in front of the two detectors. The spectral
projections of the photons was implemented by both monochromators'
slits to secure against the inhomogeneous broadening, by
effectively closing the window as it meanders. In the first set
the monochromators' slits were opened to spectral resolution of
$200\mu$eV, as independently determined by measuring the spectral
lines of a low pressure mercury lamp. In the second set the
resolution was set to $25\mu$eV. This set is shown in
Fig.~\ref{fig:two}d). The normalized integrated numbers of
coincidences from these polarization quantum tomography
measurements \cite{ref23} are then used to generate the density
matrices in the H V basis, as displayed in Fig.~\ref{fig:three}a
for the first set and Fig.~\ref{fig:three}b for the second. The
experimentally obtained density matrices was then fitted to the
form of Eq.~\eqref{eq:density_matrix}. In the first case we
obtained $|\alpha|^2=|\beta|^2=0.50\pm 0.02$ and $\gamma'=0.03\pm
0.04 + i(0.00 \pm 0.04)$. This agrees with the theoretical
estimate of $|\gamma'|=\Gamma/2\Delta\approx 0.03$. In the second
case we obtained $|\alpha|^2=|\beta|^2=0.50\pm 0.04$ and
$\gamma'=0.05\pm 0.05 +i(0.17 \pm 0.05)$ hence $|\gamma'|=0.18\pm
0.05$ is significantly different from zero. The photon pairs are
therefore entangled with confidence level greater than 3 standard
deviations of the measurement uncertainty. Substituting
$|\gamma'|$ in Eq.~\eqref{eq:bell_inequality} gives $2.13\pm0.07$
which violates Bell's inequality.

\begin{figure}[htbp]
\epsfxsize=.38\textwidth \centerline{\epsffile{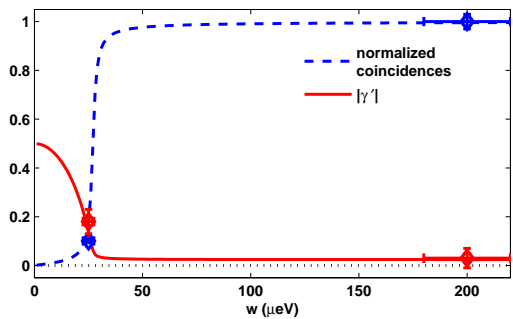}}
\vspace*{-0.2cm} \caption{Measured (symbols) and calculated
(lines) normalized number of coincidences $|P|\psi\9|^2$ and off
diagonal density matrix element $|\gamma'|$ vs. the spectral
window width . For the calculations we used $\Gamma=1.6\, \mu$ev
and $\Delta=27\,\mu$eV as experimentally determined (see text).}
\label{fig:four}
\end{figure}

In Fig.~\ref{fig:four} we plot the calculated values of the
normalized projected coincidences $|P|\psi\9|^2$  and $|\gamma'|$
as a function of the spectral window width $w$. The measured data
points at $25\mu$eV and $200\mu$eV agree to within the error bars
with the calculations. This confirms that $|\6 d_H |d_V \9|
\approx 1$. Thereby, we have verified that no ``which path''
information is contained in the final state of the SCQD.

The entangled photons generated in this work are not event ready
for two reasons: First, we used continuous excitation and second,
the erasure introduces randomness. However, neither is
fundamental. The excitation can be triggered on demand and the
randomness can be overcome by spectrally monitoring the discarded
photons without demolishing the entangled pair. Reducing the
detuning and increasing the radiative width is yet another
approach. Indeed, QDs with significantly smaller detuning than
ours have been reported \cite{ref26} and, at the same time, a
substantial increase in the radiative width can be achieved in
SCQD through the Purcell effect \cite{ref15,ref30}.

In summary, we demonstrated that the biexciton - exciton radiative
cascade in single semiconductor quantum dots can be used as a
source for entangled photon-pairs. By applying the Peres criterion
to the quantum tomographic data we prove that the polarization
state is entangled. Moreover, it violates Bell's inequality. The
measured degree of entanglement agrees with the calculated
radiative amplitudes. This proves that the which path information
residing in the photons color can be erased, and, at the same
time, that there is no ``which path'' information left in the
quantum dot.

We acknowledge support by the Israel Science and by the US-Israel
Binational Science Foundations.

\end{document}